\documentstyle[sprocl]{article}
\def\Journal#1#2#3#4{{#1} {\bf #2}, #3 (#4)}
\def\NPB{{\em Nucl. Phys.} B}
\def\PLB{{\em Phys. Lett.}B}
\def\PRL{\em Phys. Rev. Lett.}
\def\PRD{{\em Phys. Rev.} D}
\def\NCA{\em Nuovo Cimento}

\newcommand{\eqn}[1]{Eq.~\ref{#1}}
\newcommand{\ft}[2]{{\textstyle\frac{#1}{#2}}}
\def\diag{{\rm diag}}
\newcommand{\dr}{\raise.3ex\hbox{$\stackrel{\leftarrow}{\partial }$}{}}
\newcommand{\delr}{\raise.3ex\hbox{$\stackrel{\leftarrow}{\delta }$}{}}
\newcommand{\dsl}{\not\!\partial}
\renewcommand{\u}[1]{{\bar{#1}}}

\newcommand{\bitem}[1]{\vspace{-2mm}
\begin{#1} \setlength{\itemsep}{-5pt} }
\newcommand{\eitem}[1]{\end{#1}\vspace{-2mm}}
\def\Tr{{\rm Tr}\hskip 1pt}
\newsavebox{\uuunit}
\sbox{\uuunit}
{\setlength{\unitlength}{0.825em}
 \begin{picture}(0.6,0.7)
\thinlines
\put(0,0){\line(1,0){0.5}}
\put(0.15,0){\line(0,1){0.7}}
 \put(0.35,0){\line(0,1){0.8}}
\multiput(0.3,0.8)(-0.04,-0.02){12}{\rule{0.5pt}{0.5pt}}
\end {picture}}
\newcommand {\unity}{\mathord{\!\usebox{\uuunit}}}
\newcommand  {\Rbar} {{\mbox{\rm$\mbox{I}\!\mbox{R}$}}}
\newcommand  {\Hbar} {{\mbox{\rm$\mbox{I}\!\mbox{H}$}}}
\newcommand {\Cbar}
    {\mathord{\setlength{\unitlength}{1em}
     \begin{picture}(0.6,0.7)(-0.1,0)
        \put(-0.1,0){\rm C}
        \thicklines
        \put(0.2,0.05){\line(0,1){0.55}}
     \end {picture}}}

\newcommand{\rsemisum}{{\mathinner{\in\mkern-11mu\rule{0.15mm}
   {2.0mm}\mkern10mu}}}
\begin{document}
\thispagestyle{empty}
\begin{flushright}
KUL-TF-98/54\\
SU-ITP-98/67\\
\hfill{hep-th/9812066}\\
\today\\
\end{flushright}
\vskip3cm
\begin{center}
{\large {\bf  Conformal Symmetry  on \\[2pt] World Volumes of
Branes}$^\dagger$}
\vskip 1 cm
{\bf  Piet Claus$^{a}$, ~Renata Kallosh$^{b}$ and
Antoine Van Proeyen$^{a\, *}$}\\
\vskip.5cm
{\small
$^a$
Instituut voor theoretische fysica, \\
Katholieke Universiteit Leuven, B-3001 Leuven, Belgium\\
\
$^b$ Physics Department, \\
Stanford University, Stanford, CA 94305-4060, USA\\
}
\end{center}
\vskip 1.5cm
\centerline{\bf Abstract}
We show how the anti-de Sitter isometries of a brane solution of
supergravity theory produce superconformal invariance of their
world-volume action.  In this way linear as well as non-linear
superconformal actions are obtained in various dimensions. Two
particular examples are a conformal action with the antisymmetric tensor
in 6 dimensions in Pasti-Sorokin-Tonin formulation, and superconformal
mechanics.
\vfill
{\vfill\leftline{}\vfill
\footnoterule
\noindent
{\footnotesize
$\phantom{a}^\dagger$ Based on talks given by R.~Kallosh and A.~Van Proeyen
at the {\sl Trieste Conference on Superfivebranes and Physics in 5+1
dimensions}, Trieste, April 1--3, 1998, to appear in the proceedings.
\newline
\noindent
$\phantom{b}^*$ Onderzoeksdirecteur FWO, Belgium}  }
\newpage
\setcounter{page}{1}
\title{
CONFORMAL SYMMETRY  ON \\ WORLD VOLUMES OF   BRANES  }
\author{  P. CLAUS $^*$,  R. KALLOSH $^+$,  A. VAN PROEYEN $^*$}
\address{$^*$ Instituut voor theoretische fysica, \\
Katholieke Universiteit Leuven, B-3001 Leuven, Belgium }
\address{$^+$
 Physics Department, \\ Stanford
University, Stanford, CA 94305-4060, USA }

\maketitle
\abstracts{ We show how the anti-de Sitter isometries of a brane solution of
supergravity theory produce superconformal invariance of their
world-volume action.  In this way linear as well as non-linear
superconformal actions are obtained in various dimensions. Two
particular examples are a conformal action with the antisymmetric tensor
in 6 dimensions in Pasti-Sorokin-Tonin formulation, and superconformal mechanics. }

\section{Introduction}

The world-volume theory of certain branes is superconformal
invariant. This can be either the case for a brane living in a flat
background, or in the background produced by a stack of other
branes. In the former case the linearised action is superconformal, why in the latter
one, on
which we will concentrate for the main part of the talk,
this leads to a non-linear superconformal action.
\par
The starting point is a classical solution of supergravity describing
N coincident branes.
This  solution has a limit `the near-horizon limit' in which the
space-time metric is an anti-de Sitter (adS) geometry. (The other limit is
 flat space which leads to the linear conformal action). The
Killing vectors and Killing spinors of that geometry define a
supergroup $g$, which by itself could serve as an alternative starting
point of the construction. In that construction the geometry of a
supercoset having $g$ as its isometry group delivers all the input
for the world-volume action. The isometries determine rigid symmetries of that
world-volume action, which has further local symmetries: general
coordinate transformations and $\kappa$-symmetry. After gauge fixing of
these local symmetries, the rigid symmetries change, because the
gauge-fixing is invariant under a combination of the former rigid
symmetries with some parts of the general coordinate transformations
and $\kappa$-symmetry. This results in transformations which we
recognize as the superconformal transformations in the world-volume
theory.
\par
The above mechanism is applicable for D3 branes in 10 dimensional string theory,
M2 and M5 branes in 11 dimensions. The latter case leads to a conformal
version of the 6-dimensional antisymmetric tensor multiplet. These
are not the only examples. One may also consider intersections of
branes or other constructions in lower dimensions. One more application
which we will illustrate, is the construction in this way of a `relativistic superconformal
mechanics'. That one is obtained starting from black hole solutions
in $d=4$, $N=2$ supergravity.
\par
In Sec.~\ref{ss:algebras} we review the classification of super-anti-de
Sitter and conformal algebras. The $adS$ geometry appears in limits
of the geometry of branes (see Sec.~\ref{ss:adS}) and provides rigid symmetries
in the world-volume theory, as shown in
Sec.~\ref{ss:wv}. In that section we show how in the bosonic case
conformal symmetry arises. The same general
principles are used in Sec.~\ref{ss:scm} to obtain `relativistic superconformal
mechanics' near the horizon of a Reissner-Nordstr\"om black hole.
\par
The supersymmetric world-volume theory depends on superspace quantities. We
therefore first give some results on the supergeometry in
Sec.~\ref{ss:supergeometry}. Then we present the action
in Sec.~\ref{ss:susyth}, where
we use mainly the black hole example to illustrate the rigid
$adS$-supersymmetries, the role of $\kappa$-symmetry, and its gauge fixing.
Sec.~\ref{ss:lin} explains the
construction of the linear conformal theory, before presenting the conclusions
in Sec.~\ref{ss:conclusions}.
This talk summarizes our papers \cite{m5tens,conffads,bhsconfm} and
some unfinished work on the supersymmetrisation of the conformal
mechanics.
\section{adS and conformal superalgebras}
\label{ss:algebras}
The anti-de Sitter algebra is the algebra of Lorentz rotations $M_{\mu\nu}$
and translations $P_\mu$, where the latter do not commute:
\begin{eqnarray}
&&\left[M_{\mu\nu},M_{\rho\sigma}\right]
=\eta_{\mu[\rho}M_{\sigma]\nu}
-\eta_{\nu[\rho}M_{\sigma]\mu}\ ;\qquad
\left[P_\mu , M_{\nu\rho}\right]=\eta_{\mu[\nu}P_{\rho]}\nonumber\\
&&\left[ P_\mu,P_\nu\right]= \frac{1}{2 R^2}M_{\mu\nu}\ .
\label{adSalgebra}
\end{eqnarray}
With the opposite sign for the last commutator we would have the de Sitter algebra.
Defining $M_{d\mu}=-M_{\mu d}= R\, P_\mu$ we have generators
$M_{\hat\mu\hat \nu}=-M_{\hat\nu\hat \mu}$ with $\hat\mu=0,\ldots
,d$, and defining the metric to be $\eta_{\hat\mu\hat \nu}=\diag
(-+\ldots +-)$ the algebra can be concisely written as
\begin{equation}
\left[M_{\hat\mu\hat\nu},M_{\hat\rho\hat\sigma}\right]
=\eta_{\hat\mu[\hat\rho}M_{\hat\sigma]\hat\nu}
-\eta_{\hat\nu[\hat\rho}M_{\hat\sigma]\hat\mu} \ ,
\end{equation}
i.e. it is the algebra $SO(d-1,2)$. Note that every point is
invariant under the rotations around this point, while not invariant
under the translations. In this sense we can write
\begin{equation}
adS_d= \frac{SO(d-1,2)}{SO(d-1,1)}\ .
\end{equation}
\par
The conformal algebra, is according to the Coleman--Mandula theorem the largest
space--time symmetry which we can impose, allowing non-trivial scattering of
particles. Its usefulness is most outspoken in 2 dimensions, because then the
group is infinite dimensional.
Conformal symmetry is defined as the symmetry which preserves angles. Therefore
it should contain the transformations which change the metric up to a
factor. That implies that the symmetries are determined by the
solutions to the `conformal Killing equation'
\begin{equation}
\partial_{(\mu}\xi_{\nu)}-\ft1d \eta_{\mu\nu}\partial_\rho \xi^\rho=0\ .
\end{equation}
In dimensions $d>2$ the conformal algebra is finite-dimensional. Indeed, the
solutions are
\begin{equation}
\xi^\mu(x)=a^\mu +\lambda_M^{\mu\nu}x_\nu+\lambda_D x^\mu
+(x^2\Lambda_K^\mu-2x^\mu x\cdot \Lambda_K) \ . \label{ximu}
\end{equation}
Corresponding to the parameters $a^\mu$ are the translations $P_\mu$,
to $\lambda_M^{\mu\nu}$ correspond the Lorentz rotations $M_{\mu\nu}$, to
$\lambda_D$ are associated dilatations $D$, and $\Lambda_K^\mu$ are parameters
of `special conformal transformations' $K_\mu$. This is expressed as follows
for the full set of conformal transformations $\delta_C$:
\begin{equation}
\delta_C= a^\mu  P_\mu + \lambda_M^{\mu\nu}M_{\mu\nu}+\lambda_D D +
\Lambda_K^\mu K_\mu   \ .
\end{equation}
With these transformations, one can obtain the algebra with as non-zero commutators
\begin{equation}
\begin{array}{ll}
[M_{\mu\nu} , M^{\rho\sigma}]=
-2\delta_{[\mu}^{[\rho} M_{\nu]}{}^{\sigma]} \ ;\qquad   &
\ [P_{\mu} , M_{\nu\rho}\ ] =\eta_{\mu[\nu} P_{\rho]}\nonumber\\
\ [K_{\mu} , M_{\nu\rho}\ ] = \eta_{\mu[\nu}K_{\rho]} \ ;\qquad  &
\ [P_{\mu} , K_{\nu}\ ]= 2 (\eta_{\mu\nu} D + 2 M_{\mu\nu}) \ ,
 \nonumber\\
\ [D , P_{\mu} \ ]= P_\mu \ ;\qquad  &
\ [D , K_{\mu} \ ]=-K_\mu \ .     \label{confalg}
\end{array}
\end{equation}
This is the $SO(d,2)$ algebra. Indeed one can define
\begin{equation}
M^{\hat \mu\hat \nu}=\pmatrix{M^{\mu\nu}&\ft14(P^\mu-K^\mu)&
\ft14(P^\mu+K^\mu)\cr
-\ft14(P^\nu-K^\nu)&0&-\ft12 D\cr
-\ft14(P^\nu+K^\nu) &\ft12 D &0  }
\end{equation}
preserving the metric $\eta={\rm diag~}(-1,1,...,1,-1)$. Note that
this is the same as the anti-de Sitter algebra in $d+1$ dimensions
\begin{equation}
Conf_{d} = adS_{d+1}\ ,
\end{equation}
which is the basic observation for the adS/CFT correspondence.
\par
In general, fields $\phi^i(x)$  have the following
transformations under the conformal group:
\begin{eqnarray}
\delta_C \phi^i(x)&=& \xi^\mu(x)\partial_\mu \phi^i(x)
+ \Lambda_M^{\mu\nu}(x)\, m_{\mu\nu}{}^i{}_{j}\phi^j(x)
\nonumber\\ &&+ w_i\,\Lambda_D(x)
\,\phi^i(x)+ \Lambda_K^\mu\left( k_\mu \phi\right) ^i(x)\ ,
\label{deltaC}
\end{eqnarray}
where the $x$-dependent rotation $\Lambda_{M\,\mu\nu}(x)$ and $x$-dependent
dilatation $\Lambda_D(x)$ are given by
\begin{eqnarray}
\Lambda_{M\,\mu\nu}(x)&=&\partial_{[\nu}\xi_{\mu]}=\lambda_{M\,\mu\nu}
-4x_{[\mu} \Lambda_{K\,\nu]} \ , \nonumber\\
\Lambda_D(x)&=&\ft1d \partial_\rho \xi^\rho =
\lambda_D -2 x\cdot \Lambda_K  \ .   \label{Lambdax}
\end{eqnarray}
The matrix $m_{\mu\nu}$ represent the usual Lorentz rotations, and the number
$w_i$ is called the Weyl weight of the field. Any term in the action should
have Weyl weight $d$, counting 1 for a derivative. But furthermore there is one
more requirement for special conformal transformations:
\begin{equation}
\delta_K S=2\Lambda_K^\mu  \int d^dx\,
\frac{{\cal L}\dr}{\partial(\partial_\nu\phi^i)}
\left(- \eta_{\mu\nu}w_i\phi^i +2m_{\mu\nu}{}^i{}_j\phi^j \right)
+\Lambda_K^\mu\frac{ S\delr}{\delta \phi^i(x)}(k_\mu\phi)^i(x)\ .
\label{deltaLK}
\end{equation}
where $\dr$ indicates a right derivative. The first terms originate
from the $K$-transformations contained in \eqn{ximu} and \eqn{Lambdax}.
In most cases these are sufficient to find the invariance and no
$(k_\mu\phi)$ are necessary. However, e.g. for the theories which one obtains
from $adS$ backgrounds, such extra terms are present which depend on the field $r$,
the distance from the brane.
\par
In supersymmetric theories, the conformal symmetry implies the presence of a
second supersymmetry $S$, usually denoted as `special supersymmetry'. Indeed,
the commutator of the special conformal transformations, and the ordinary
supersymmetry $Q$ implies this $S$ due to $[K_\mu,Q]=\gamma_\mu S$.
\par
\begin{table}[t]\caption{Lie superalgebras of classical type. For the
real forms of $SU(m|m)$, the one-dimensional subalgebra of the bosonic
algebra is not part of the irreducible algebra. Furthermore, in that case
there are subalgebras obtained from projection of those mentioned here
with only one factor $SU(n)$, $S\ell (n)$, $SU^*(n)$ or $SU(n-p,p)$ as
bosonic algebra.
\label{tbl:LieSA}}
\begin{center}\begin{tabular}{|l|l|l|l|l|}\hline
Name & Range  & Bosonic algebra  & Defining& Number of  \\
     &        &                  & repres. & generators\\ \hline \hline
$SU(m|n)$&$m\geq 2$&$ SU(m)\oplus SU(n)$&$(m,\bar n)\oplus
$&$m^2+n^2-1,$\\
     & $ m\neq n $       & \ \ \ $\oplus U(1)$&\ $(\bar m,n)$ &\ \ \  $2mn  $\\
& $ m= n $&\ \ \ no $U(1)$   & 
&$2(m^2-1),2m^2$\\
     \hline
 \multicolumn{2}{c|}{
\begin{tabular}{|l}
$S\ell(m|n) $       \\
$SU(m-p,p|n-q,q) $   \\
$SU^*(2m|2n) $\\
$S\ell \,{}^\prime(n|n)$
\end{tabular}}
& \multicolumn{3}{l|}{\begin{tabular}{l}
$\begin{array}{l}
S\ell (m)\oplus S\ell (n)  \\
SU (m-p,p)\oplus SU(n-q,q)\\
SU^*(2m)\oplus SU^*(2n)
\end{array}  $
\hspace{-15pt} $\left.
\begin{array}{l}
\oplus SO(1,1)   \\
\oplus U(1)   \\
\oplus SO(1,1)
\end{array} \hspace{-5pt}\right\}$\hspace{-10pt} \begin{tabular}{l}
if  \\
 $m\neq n$
\end{tabular}
\\ $S\ell (n,\Cbar)$\end{tabular}}
\\ \hline\hline
$OSp(m|n)$&$ m\geq 1$ & $SO(m)\oplus Sp(n)$ & $(m,n)$ & $\ft12(m^2-m+$\\
          & $ n=2,4,..$ &                   &         &\ \ \  $n^2+n),mn$\\ \hline
 \multicolumn{2}{c|}{
\begin{tabular}{|l}
\phantom{$SU(m-p,p|n-q,q) $}  \\[-4mm]
$OSp(m-p,p|n)$  \\
$OSp(m^*|n-q,q)$
\end{tabular} }&
 \multicolumn{2}{c}{
\begin{tabular}{l}
\phantom{$SU (m-p,p)\oplus SU(n-q,q)$}\hspace{-5pt}  \\[-4mm]
$ SO(m-p,p)\oplus Sp(n)$ \\
  $ SO^*(m)\oplus USp(n-q,q)$
\end{tabular} }&
\begin{tabular}{l} $n$ even
 \\ $m,n,q$ even
\end{tabular}
\\ \hline\hline
$D(2,1,\alpha)$ & $0<\alpha\leq 1$ & $SO(4)\oplus S\ell (2)$&$(2,2,2)$&9, 8 \\
                \hline
 \multicolumn{2}{c|}{
\begin{tabular}{|l}
\phantom{$SU(m-p,p|n-q,q) $}   \\[-4mm]
$D^p(2,1,\alpha)$
\end{tabular} }&
 \multicolumn{2}{c}{
\begin{tabular}{l}
\phantom{$SU (m-p,p)\oplus SU(n-q,q)$} \\[-4mm]
 $SO(4-p,p) \oplus S\ell (2)$
\end{tabular} }&$p=0,1,2$ \\ \hline\hline
$F(4)$ & & $\overline {SO(7)}\oplus S\ell (2)$ & $(8,2)$& 21, 16 \\ \hline
 \multicolumn{2}{c|}{
\begin{tabular}{|l}
\phantom{$SU(m-p,p|n-q,q) $} \\[-4mm]
$F^p(4)$\\ $F^p(4)$
\end{tabular} }&
 \multicolumn{2}{l}{
\begin{tabular}{l}
\phantom{$SU (m-p,p)\oplus SU(n-q,q)$} \\[-4mm]
$SO(7-p,p)\oplus S\ell (2)$\\ $SO(7-p,p)\oplus SU(2)$
\end{tabular} }&
\begin{tabular}{l}
$p=0,3$    \\ $p=1,2$
\end{tabular}
\\ \hline\hline
$G(3)$ & & $G_2\oplus S\ell (2)$ & $(7,2)$ & 14, 14\\  \hline
 \multicolumn{2}{c|}{
\begin{tabular}{|l}
\phantom{$SU(m-p,p|n-q,q) $} \\[-4mm]
$G_p(3)$
\end{tabular} }&
 \multicolumn{2}{c}{
\begin{tabular}{l}
\phantom{$SU (m-p,p)\oplus SU(n-q,q)$} \\[-4mm]
$G_{2,p}\oplus  S\ell (2)$
\end{tabular}} & $p=-14,2$\\ \hline\hline
$P(m-1)$&$m\geq 3$ & $S\ell (m)$ & $(m\otimes m)$&$m^2-1,m^2$ \\ \hline\hline
$Q(m-1)$&$m\geq 3$ & $SU(m)$ & Adjoint &$m^2-1,m^2-1$ \\ \hline
 \multicolumn{2}{c|}{
\begin{tabular}{|l}
\phantom{$SU(m-p,p|n-q,q) $} \\[-4mm]
$Q(m-1)$ \\
$Q((m-1)^*)$  \\
$UQ(p,m-1-p)$
\end{tabular} }&
 \multicolumn{2}{c}{
\begin{tabular}{l}
\phantom{$SU (m-p,p)\oplus SU(n-q,q)$} \\[-4mm]
$S\ell (m)$\\
 $SU^*(m)$ \\
 $SU(p,m-p)$
\end{tabular}} & \\ \hline\hline
\end{tabular}\end{center} \end{table}
\par
The anticommutator $\left\{Q,S\right\}$ generates also an extra
bosonic algebra (sometimes called $R$ symmetry). The whole superalgebra
can be represented in a supermatrix, as e.g. (symbolically)
\begin{equation}
\pmatrix{SO(d,2)& Q+S\cr Q-S&R}\ .
\end{equation}
We can consider such superalgebras in general. That is what Nahm \cite{nahm}
did in his classification. The requirements for a superconformal algebra
in $d$ or a super-adS algebra in $d+1$ are:
\bitem{enumerate}
\item $SO(d,2)$  should appear as a factored subgroup of the
bosonic part of the superalgebra. For Nahm, this requirement was
motivated by the Coleman-Mandula theorem, but here it is imposed in order to
have that the bosonic algebra is the isometry algebra of a space
which has the $adS$ space as a factor.
\item  fermionic generators should
sit in a spinorial representation of that group.
\eitem{enumerate}
One can then consider the list \cite{LieSA} of simple Lie superalgebras\footnote{Technically
speaking this is the list of Lie superalgebras of classical type, i.e.
those for which the fermionic generators are in an irreducible or
completely reducible representation of the bosonic algebra. For more
explanations, see e.g. the review \cite{Annecy}.} in Table~\ref{tbl:LieSA}.
In this table\footnote{The table has been changed after the first version
of this paper, correcting also the table in \cite{Annecy}. These
corrections have been found in discussions with S.~Ferrara.}, `defining
representation' gives the  fermionic generators as a
representation of the bosonic subalgebra. The `number of generators' gives the
numbers of (bosonic,fermionic) generators in the superalgebra. We mention first
the algebra as an algebra over $\Cbar$, and then give different real forms of
these algebras\footnote{The conventions which we use for groups is that
$Sp(2n)=Sp(2n,\Rbar)$ (always even entry), and $USp(2m,2n)=U(m,n,\Hbar)$.
$S\ell (n)$ is $S\ell (n,\Rbar)$, and thus $Sp(2)=S\ell (2)=SU(1,1)$. Further,
$SU^*(2n)=S\ell (n,\Hbar)$ and $SO^*(2n)=O(n,\Hbar)$. The exceptional group
$G_2$ has only 2 real forms, the compact one $G_{2,-14}$ and the `normal form'
$G_{2,2}$.}.
The names which we use for the real forms \cite{realLieSA} is for some different from those in the
mathematical literature, and chosen such that it is most suggestive of its
bosonic content. There are isomorphisms as $SU(2|1)=OSp(2^*|2,0)$,
and $SU(1,1|1)=S\ell (2|1)=OSp(2|2)$. In the algebra $D(2,1,\alpha)$
the three $S\ell (2)$ factors of the bosonic group in the anticommutator of
the fermionic generators appear with relative weights $1$, $\alpha$
and $-1-\alpha$. The real forms contain respectively
$SO(4)=SU(2)\times SU(2)$, $SO(3,1)=S\ell (2,\Cbar)$ and
$SO(2,2)=S\ell (2)\times S\ell (2)$. In the first and last case
$\alpha$ should be real, while $\alpha= 1 +ia$ with real $a$
for $p=1$. In the
limit $\alpha=1$ one has the isomorphisms $D^p(2,1,1)=OSp(4-p,p|2)$.
\par
Scanning through that list, one finds those algebras which satisfy the
conditions for super-$adS$ or superconformal algebras.
The result are algebras
with maximal $d=6$\footnote{Note the particular case of $d=6$, where we use the notation
$OSp(8^*|N)$ for the superconformal algebra. Often, including previous articles of ourselves,
it was written as $OSp(6,2|4)$, not paying attention to the existing real forms.
In fact, in the series $OSp(m,n|2p)$ the algebra $Sp(2p)$ is non-compact. The possibility for
a compact $R$-symmetry group $USp(2p)$ exists due to the isomorphism $SO^*(8)=SO(6,2)$, and thus
works only for this signature.}. The result\footnote{We mention the superalgebra with
compact $R$-symmetry group.} is given in
Table~\ref{tbl:sca}, except for $d=2$.
For $d=2$ the finite bosonic adS or conformal algebra is $SO(2,2)\approx
SO(2,1)\oplus SO(2,1)$, i.e. the sum of two $d=1$ algebras. The super-adS or
superconformal algebra is then the sum of two $d=1$ algebras of
Table~\ref{tbl:sca}.
\begin{table}[t]\caption{Super $adS_{d+1}$ or $conf_d$ algebras.
\label{tbl:sca}}\vspace{0.4cm}
\begin{center}\begin{tabular}{|lllc|}\hline
$d$& superalgebra& $R$ & number of fermionic\\ \hline
$1$&  $OSp(N|2)$ & $O(N)$    & $2N$ \\
   &  $SU(N|1,1)$  &$SU(N)\times U(1)$ for $N\neq 2$ &$ 4N$ \\
   &  $SU(2|1,1)           $    &$SU(2)              $ &$ 8   $\\
   &  $OSp(4^*|2N)         $    &$SU(2)\times USp(2N)$ &$ 8N  $\\
   &  $G(3)                $    &$G_2                $ &$ 14  $\\
   &  $F^0(4)                $    &$SO(7)              $ &$ 16  $\\
   &  $D^0(2,1,\alpha)     $    &$SU(2)\times SU(2)  $ &$  8  $\\  \hline
$3$&$ OSp(N|4)   $ &$ SO(N) $&$ 4N $\\   \hline
$4$&$ SU(2,2|N)  $ &$SU(N)\times U(1)$ for $N\neq 4$&$ 8N$\\
   &$ SU(2,2|4)  $ &$  SU(4)$ & 32\\  \hline
$5$&$ F^2(4)       $ &$ SU(2) $ & 16 \\       \hline
$6$&$ OSp(8^*|N) $ &$  USp(N)\ \ (N $ even)& $8N$ \\
\hline
\end{tabular}
\end{center}
\end{table}
Notice that these are the finite part of infinite dimensional superconformal
algebras in 2 dimensions. For a classification of the infinite
superconformal algebras, see \cite{classinf}. One may
also relax the condition that the bosonic algebra contains the algebra
$SO(d,2)$ as a factored subgroup of the whole bosonic algebra, and suffice with
having it as some subgroup. Then the other bosonic symmetries are not
necessarily scalars and the Coleman-Mandula theorem is violated. However, this
may still be relevant where branes are present and has been used e.g. in
\cite{JWvHAVP} to propose the $OSp(1|32)$ as super $adS_{11}$ or $conf_{10}$.
In that case one has
\begin{equation}
\left[ Q,Q\right]=\Gamma^\mu P_\mu  + \Gamma^{\mu \nu} Z^{(2)}_{\mu \nu}
+\Gamma^{\mu \nu\rho\sigma\tau}Z^{(5)}_{\mu \nu\rho\sigma\tau}   \ .
\end{equation}
This algebra is now known as the $M$-theory algebra \cite{Mfromsa}.

\section{adS geometry from a hypersurface and from branes}
\label{ss:adS}
To obtain a space with $adS_d$ metric, we start from defining it as
a submanifold of a $(d+1)$-dimensional space with a flat metric of
$(d-1,2)$ signature
(for convenience we taken here $\mu=0,\ldots ,d-2$)
\begin{equation} \tabcolsep 1pt
\begin{array} {ccccc}
ds^2=&dX^\mu \eta_{\mu\nu} dX^\nu&-&dX^+dX^- &      \\
     & (d-2,1)                     &+&(1,1)    &\Rightarrow (d-1,2)
\end{array}\tabcolsep 6pt
\end{equation}
The $adS$ space is the submanifold determined by the equation (again
$SO(d-1,2)$ invariant)
\begin{equation}
X^\mu \eta_{\mu\nu} X^\nu-X^+X^- +R^2 =0\ .  \label{adSsubmanif}
\end{equation}
On the hypersurface one can take several sets of coordinates. E.g. the
horospherical coordinates $\{x^\mu,z \}$ by
\begin{eqnarray}
X^-&=&z^{-1}\nonumber\\
X^\mu &=& z^{-1}  x^\mu\nonumber\\
X^+&=&\frac{x_\mu^2+R^2  z^2 }{z }  \ .
\end{eqnarray}
The latter being the solution of \eqn{adSsubmanif} given the first two.
The induced metric on the hypersurface is
\begin{equation}
ds^2=\frac{1}{z^2}\left(  dx_\mu^2+R^2 dz ^2 \right) \ . \label{adSmetric}
\end{equation}
The $SO(d-1,2)$ is linearly realized in the embedding $(d+1)$-dimensional
space, and these transformations, ($\hat\mu=\mu,+,-$ and
$\Lambda^{\hat\mu\hat\nu}=-\Lambda^{\hat\nu\hat\mu}$)
\begin{equation}
\delta X^{\hat\mu}=\Lambda^{\hat\nu\hat\rho}M_{\hat\nu\hat\rho}
X^{\hat\mu}=-\Lambda^{\hat\mu}{}_{\hat\nu} X^{\hat\nu}\ .
\end{equation}
are on the $adS$ space distorted to
\begin{eqnarray}
\delta_{adS} x^\mu&=& -\hat \Lambda^\mu{}_- -\Lambda^{\mu\nu}x_\nu
-\Lambda^+{}_+ x^\mu \nonumber\\ &&
-(x^2+R^2 z^2)\Lambda^\mu{}_++2x^\mu x_\nu \Lambda^\nu{}_+  \nonumber\\
\delta_{adS} z&=&-z\left( \Lambda^+{}_+ -2x_\mu\Lambda^\mu{}_+
\right)\ .   \label{adSiso}
\end{eqnarray}
\par
On the other hand, the $p$-brane solutions of $D$-dimensional supergravity
have a metric of the form
\begin{eqnarray}
ds^2&=&H_{\rm brane} ^{-{2\over d}}  dX_m^2 + H_{\rm brane}
^{2\over \tilde d} dX_{m'}^2 \ ;\qquad d=p+1\ ;\ \tilde d\equiv D-p-3 \nonumber\\
 H_{\rm brane} &=& 1+ \left ({R\over r}\right )^{\tilde d}\ ;\qquad
r^2=X^{m'} X^{m'}\ ,
\end{eqnarray}
where $m= 0,1, \dots p$ denotes the directions along the brane,
and $m' =1, \dots D-p-1$ those orthogonal to the brane. The solutions
describe $N$ coincident branes, and the parameter $R$ is proportional to
$N^{1/\tilde d}$. When the constant 1 is neglected in the expression for
the harmonic function in the
transverse coordinates $H$, one obtains an $adS_{d+1}$ metric. Indeed then
\begin{equation}
ds^2_{hor}=\left ({r\over R}\right ) ^{2\tilde d\over d}  dX_m^2 +
\left ({R\over r}\right )^2  dr^2+ R^2d^2\Omega\ , \label{ds2hor}
\end{equation}
and identifying $z$ in \eqn{adSmetric} with
$\left({R\over r}\right)^{\tilde d\over d}$ we find that $(X^m,r)$ is a
$(d+1)$-dimensional adS space and the remainder is a $(\tilde d+1)$-sphere.
\par
The described limit can be seen in 3 different ways:
\bitem{itemize}
\item  We can see the brane solution as an interpolation \cite{GT} between the
asymptotically flat region where $H= 1$, and a
near horizon anti-de Sitter geometry where
$H= \left ({R\over r}\right )^{\tilde d}$.
\item The limit can also be seen as a large N (many branes solution) limit
\cite{Maldacena}. This will correspond in the field theory to large N  for the
$SU(N)$ gauge theory.
\item there is a special duality transformation that removes the
constant \cite{specialdual}.
\eitem{itemize}
This mechanism applies in various situations. We give in Table~\ref{tbl:confpbr}
\begin{table}[t]\caption{Brane solutions with horizon geometry $adS_{d+1}
\times S^{\tilde d+1}$, where $d=p+1$. The number
$w=d/{\tilde d}$ is the Weyl weight of the scalars in the conformal
theory.
\label{tbl:confpbr}}\vspace{0.4cm}
\begin{center}
\begin{tabular}{|l|c|ccc|l|}\hline
                              & $D$ & $d$&$\tilde d$ & $w$ &superalgebra \\ \hline
M5                            & 11& 6&3 & 2       &$ OSp(8^*|4)$ \\
M2                            & 11& 3&6 & $\ft12$ &$OSp(8|4)$ \\
D3                            & 10& 4&4 & 1       &$SU(2,2|4)$ \\
Self-dual string (D1+D5)      & 6 & 2&2 & 1       &$SU(1,1|2)\oplus SU(1,1|2)$ \\
Magnetic string               & 5 & 2&1 & 2       & $SU(1,1|2)\oplus SU(1,1)$\\
Tangerlini black hole         & 5 & 1&2 & $\ft12$ & $SU(1,1|2)\,\rsemisum SU(2)$\\
Reissner-Nordstr\"om black hole& 4& 1&1 & 1     &$SU(1,1|2)$\\
\hline
\end{tabular}
\end{center}
\end{table}
cases where the manifold is always of the form $adS_{d+1}\times S^{\tilde d+1}$.
The isometry groups are thus $SO(d,2)\times SO(\tilde d+2)$. For later
convenience we mention also $w\equiv d/\tilde d$. E.g. the
self-dual string can also be obtained from compactifying a
10-dimensional string theory on $adS_3\times S^3 \times E_4$, where
$E_4$ denotes an Euclidean space. The Tangerlini black hole has been discussed, together
with related rotating black holes in 5 dimensions in \cite{GMT},
where the mentioned superalgebra was obtained (the extra $SU(2)$ rotates
the supercharges, it is in fact $D^2(2,1,0)$). We are not aware of a
similar result for the magnetic string, but we conjecture the
appearance of the superalgebra in the table.
In many cases
other manifolds than simple spheres appear in the compactification,
see e.g.  \cite{KPSk2}.

\section{Bosonic world-volume theory}
\label{ss:wv}

The world-volume theory is in general a sum of a Born-Infeld term, a Wess-Zumino term
and an extra one in case of the $M5$ brane. We first concentrate on
the bosonic part.
\begin{eqnarray}
S_{cl}&=&S_{BI}+S_{WZ} + S'   \nonumber\\
S_{BI}&=&- \int d^{d}  \sigma  \sqrt{- \det \left( g^{\rm ind}_{\mu\nu}   +T_{\mu\nu}
\right) }
\ ;\hspace{4mm}g^{\rm ind}_{\mu\nu}=  \partial_\mu X^{ M} \partial_\nu X^{ N}G _{ M
N}\,,
\label{WVactions}
\end{eqnarray}
where $G_{MN}$ is the metric on target space, solution of the
supergravity theory. We use for the latter immediately the near-horizon form
\eqn{ds2hor}. Concentrating on the first three and the
last case of Table~\ref{tbl:confpbr},
\begin{eqnarray}
RN \mbox{ or } M2
&\qquad T_{\mu\nu}=0&\qquad S'=0\ ;\nonumber\\
D3&\qquad T_{\mu\nu}=F_{\mu\nu}&\qquad S'=0\ ;\nonumber\\
M5&\qquad T_{\mu\nu}=i{\cal H}^*_{\mu\nu}&\qquad S'=
\int d^6\sigma\,
{\cal H}^{*\mu\nu} {\cal H}_{\mu\nu}\ .
\end{eqnarray}
In the case of $M5$ we use the Pasti-Sorokin-Tonin (PST) formulation
\cite{PST} for describing self-dual tensors. That means that there is
an  auxiliary field $a$ apart from the antisymmetric tensor $B_{\mu \nu}$, and
\begin{eqnarray}
&& u_\mu=\partial_\mu a   \ ;\qquad
{\cal H}_{\mu\nu\rho}=
3\partial_{[\mu} B_{\nu\rho ]}\ ; \nonumber\\
&&{\cal H}_{\mu\nu}=\frac{u^\rho}{\sqrt{u^2}}{\cal H}_{\mu\nu\rho}
\ ;\qquad
{\cal H}^*_{\mu\nu}=\frac{u^\rho}{\sqrt{u^2}}{\cal H}^*_{\mu\nu\rho}\ .
\label{HwithPST}
\end{eqnarray}
The field equations of $B_{\mu \nu}$ and extra gauge invariances imply that $a$
is a gauge degree of freedom and ${\cal H}_{\mu \nu\rho}$ is
self dual.
\par
The input which we thus need for the world-volume action is $G_{MN}$
and the Wess-Zumino term. The latter is related to the integral of
the $(p+2)$-field strength form of the p-brane. Its exact form can
vary for different cases, see e.g.
\begin{equation}
\begin{array}{lclcl}
M5 &:& \int \Omega_7 &;\qquad& \Omega_7= dX_0 \ldots dX_5 dH^{-1}\ ;\\
D3 &:& \int \Omega_5 + \tilde\Omega_5 &;\qquad& \Omega_5= dX_0 \ldots dX_3
dH^{-1}\ ;\\
BH &:& \int F &;\qquad& F= dX^0 dH^{-1}\ .
\end{array}
\end{equation}
\par
The isometries of the solution lead to rigid symmetries of the
action. These are thus the
$SO(\tilde d+2)$ rotations and the adS isometries
\begin{eqnarray}\label{hatxi2}
\delta_{adS}( \xi) X^m&\equiv & - \hat \xi ^m(X, r )=-\xi^m(X)-  (wR)^2 \left({R\over r }\right)^{2/w}
\Lambda_{K}^m\nonumber\\
\delta_{adS}(\xi)  X^{m'}&=&w \Lambda_D(X) X^{m'}=
w (\lambda_D - 2 X^m \Lambda_{K\,m})   X^{m'}\ ,
\end{eqnarray}
In the full theory there is also rigid supersymmetry,
determined by the Killing spinors of the metric.
\par
Note that in case of $d=1$ or 2 infinite dimensional symmetries
exist. In $adS_3$ they have been found as asymptotic geometries
by Brown and Henneaux \cite{BrownHenn}.
That means that these are not symmetries of the action, but rather
between different geometries which have $adS_3$ as near-horizon limit.
They form the Virasoro algebra of which $SO(2,2)=SU(1,1)\times SU(1,1)$
is the finite dimensional subgroup. However, it has also been found
that the action
$\int d^2\sigma \sqrt{-\det (g_{\mu\nu} + F_{\mu\nu})}$
has an infinite symmetry group \cite{BGS}. For any isometry with
$h_M(X)$ as Killing vector, the action is invariant under
\begin{eqnarray}
\delta X^M&=& h^M(X)\, \lambda({\cal F})\ ;\qquad
{\cal F}=-\frac{\epsilon^{\mu\nu}F_{\mu\nu}}{2\sqrt{g}}
\ ;\nonumber\\
\delta V_\mu &=&
-\lambda'( {\cal F}) \sqrt{g}\, (1+{\cal F}^2)\epsilon_{\mu\nu}
\left( \partial^\nu X^M\right) h_M(X) 
\ ,\end{eqnarray}
where $\lambda( {\cal F})$ is an arbitrary function which provides a
Ka\v{c}-Moody extension of the isometry group.
\par
Furthermore there are the local symmetries: world-volume diffeomorphisms and
its fermionic partner: the $\kappa$-symmetry, which we will discuss
in section~\ref{ss:susyth}.  Gauge fixing of the diffeomorphisms,
e.g. by identifying the first $d$ space-time fields with the
coordinates on the brane $X^\mu(\sigma)=\sigma^\mu$ leaves invariant
linear combinations of the rigid isometries with local symmetries
where the parameters of the latter are determined functions of the
rigid parameters of the isometries and local fields.
\begin{equation}
\delta_C(\xi)=\delta_{adS}(\xi)  +\delta_{ld}(\eta=\hat \xi)\ .
\end{equation}
The result is
that these remaining symmetries take the form of conformal
transformations on the world-volume. The remaining scalar fields are
$X^{m'}$, which are scalars of Weyl weight $w$ in Table~\ref{tbl:confpbr}.
There are extra parts in the special conformal transformations
\begin{equation} \Lambda_K^\mu k_\mu \phi= \delta_{ld}\left(
\eta^\mu=(wR)^2 \left({R\over r }\right)^{2/w}
\Lambda_{K}^\mu  \right)\ .
\end{equation}
\par

We will see below how a similar mechanism works for the fermionic
sector, and the superconformal algebra appears where the $R$ symmetry
group is provided by the isometry group of sphere.

\section{Relativistic conformal mechanics}
\label{ss:scm}
One may apply the above mechanism to get conformal
symmetry in one dimension. This is the case $p=0$ or $d=1$ and
$R=M$, the mass of a black hole solution in $d=4$, $N=2$ pure supergravity. The
solution near the horizon has the Bertotti-Robinson (BR) metric
\begin{equation}
ds^2=-\left ({r\over M}\right ) ^2  dt^2 +
\left ({M\over r}\right )^2  dr^2+ M^2d^2\Omega\ ;\qquad
W= \frac{Q}{M^2} rdt + P\cos\theta d\phi \ ,
\end{equation}
where $(r,\theta,\phi)$ are polar coordinates around the black hole.
$W_\mu$ is the graviphoton which has in this solution electric charge
$Q$ and magnetic charge $P$, satisfying $P^2+Q^2=M^2$.
The supergravity algebra of $d=4$, $N=2$  contains
\begin{eqnarray}
\left\{\bar \epsilon_1^i Q_i,\bar \epsilon_2^j Q_j\right\}&=& \bar \epsilon_2^i
\gamma^\mu \epsilon_{1i} \left(P_\mu-  \omega_\mu^{ab}M_{ab}-W_\mu G-\bar \psi_\mu ^kQ_k\right)
\nonumber\\ &&
+ \bar \epsilon^i_2 \epsilon^j_1 \varepsilon_{ij}\left( G+  F^{ab}  M_{ab}\right)
+ \bar \epsilon^i_2 \gamma_5\epsilon^j_1 \varepsilon_{ij}\tilde F^{ab}  M_{ab}\ ,
\end{eqnarray}
where $G$ is the gauge transformation of the graviphoton, whose field strength is $F$,
and $\tilde F$ is the dual of $F$. In the
solution, the gravitino $\psi_\mu $ vanishes, and
the non-zero values of spin-connection $\omega$, and gauge
field $W_\mu $ conspire to promote the solution to a BPS state.
\par
In the world line action one can introduce a WZ term with
electric charge $q$ and magnetic charge $p$ for the particle in the background
with charges $(Q,P)$
\begin{eqnarray}
S&=& m S_{BI}+ S_{WZ}\nonumber\\
 &=& -m\int d\tau\,\sqrt{- g_{00}^{ind}} +q\int_{{\cal M}_2} F
 +ip\int_{{\cal M}_2} \tilde F
 \nonumber\\
g_{00}^{ind}  &=& \left(\frac{2M}{\rho}\right) ^4\left[-\left( \dot t\right) ^2
+ \left( \frac{\rho}{2M} \dot \rho\right) ^ 2 \right] +M^2\left[
\dot \theta^2+\sin^2\theta\dot \phi^2\right]\ ,
\end{eqnarray}
where $F$ is the pullback of $dW$ on a 2-dimensional manifold ${\cal
M}_2$ which has the worldline as its boundary, and
we have used a new radial variable $\rho$ defined by
\begin{equation}
\frac{r}{M} =\left(\frac{2M}{\rho}\right) ^2 \ .
\end{equation}
We can identify the
space-time $t$ with the world-line parameter $\tau$, as gauge choice
of $\tau$-reparametrizations.
\par
For pure electric particles in an electric background, $p=P=0$,
the Hamiltonian gets the interesting form
\begin{equation}
H = \left(\frac{2M}{\rho}\right)^2 \left[ \sqrt{m^2 +
\frac{\rho^2p_\rho^2 + 4L^2}{4M^2}} -q\right]\,,
\end{equation}
where $L^2=p_\theta^2+\sin^{-2}\theta p_\phi^2$ is the angular momentum. This
Hamiltonian can be seen as $H=-p_0$ solving
\begin{equation}(p_0-qW)^2 G^{00}+ p_{m'}G^{m'n'}p_{n'}+m^2=0\ .\end{equation}
This thus describes a charged particle in the BR background. We may write also
\begin{eqnarray}
H= {p_\rho^2\over 2f} + {mg\over \rho^2 f}\ ;\quad &&
f=\ft12 \left[ \sqrt{m^2 + (\rho^2p_\rho^2 + 4L^2)/4M^2} +q\right]
\nonumber\\
&& m\,g=2M^2(m^2-q^2) + 2L^2 \ .
\end{eqnarray}
The limit of large black hole mass
\begin{equation}
M\rightarrow \infty\ ; \qquad (m-q)\rightarrow 0\ ;\qquad M^2(m-q)
\mbox{ fixed}
\end{equation}
gives $f\rightarrow m$, and
is the conformal mechanics of de~Alfaro, Fubini and Furlan \cite{DFF}.
We denote this as {`\it non-relativistic conformal mechanics'}, and the full
action with finite black hole mass as {\it relativistic conformal mechanics}.
Similarly the supersymmetric case generalises \cite{AP}, which is a
superconformal mechanics model, to a new {\it relativistic
superconformal mechanics.}
\par
The conformal invariance appears by the general mechanism.
Changing the $adS$
isometries by time-diffeomorphisms due to the gauge choice.
The $adS$ isometries (parameters $a$, $b$, $c$) and general coordinate transformation (parameter $\xi(\tau)$) act as
\begin{eqnarray}
\delta t&=& a+b\,t+c\,t^2+ c \frac{M^4}{r^2} +\xi(\tau)\dot t\nonumber\\
\delta r&=& -r \left( b+2c\, t\right)+ \xi(\tau)\dot r\ .
\end{eqnarray}
After the gauge choice $t=\tau$ the reparametrization parameter is constrained:
\begin{equation}
-\xi(\tau)=a+b\,\tau+c\,\tau^2 +c\frac{M^4}{r^2} =0 \ ,
\end{equation}
and $b$ takes the role of dilatation parameter. $r$ transforms as a scalar
with Weyl weight $w=1$.

\section{Supergeometry} \label{ss:supergeometry}
So far, we neglected the fermionic sector. For world-volume theories
of supersymmetric branes in a supersymmetric background, we need the complete
supergeometry of the background. The complete interacting
world-volume theory indeed depends on the geometric superfields of the background
to all orders in anticommuting coordinates $\theta$.
Before discussing the world-volume action in section~\ref{ss:susyth} we first
explain the two methods to obtain these
quantities in curved space: gauge completion or supergravity superspace and
supercoset methods. The method called {\em `gauge completion'} \cite{Gaugecompl}
 works in principle
for generic backgrounds.  One starts with
comparing the component expressions with the general transformations in
superspace (the superspace coordinates are $Z^\Lambda = \{X^M, \theta^A\}$),
i.e.~superdiffeomorphisms,
local lorentz transformations and superspace gauge
transformations with respectively parameters $\Xi^\Lambda$, $L^{\u{MN}})$
and $\Sigma$ (considering the case $p=0$). They act on
the fields as
\begin{eqnarray}
\Delta E_\Lambda{}^{\u\Lambda} &=& \Xi^\Pi \partial_\Pi E_\Lambda{}^{\u \Lambda}
+ (\partial_\Lambda \Xi^\Pi) E_\Pi{}^{\u\Lambda} +
E_\Lambda{}^{\u\Sigma} L_{\u \Sigma}{}^{\u \Lambda}\,,\nonumber\\
\Delta A_{\Lambda} &=&  \Xi^\Pi \partial_\Pi A_{\Lambda}
+  (\partial_{\Lambda} \Xi^\Pi) A_\Pi+ \partial_\Lambda \Sigma\ ,
\end{eqnarray}
where barred indices denote those in flat coordinates.
One compares also the algebra, first at the $\theta=0$ level,
defining
\begin{eqnarray}
&& \left.E_\Lambda{}^{\u\Lambda} \right|_{\theta=0}=\pmatrix{e_M{}^{\u M}&\psi_M{}^{\u A}\cr
0&{\cal A}_A{}^{\u A}}
\ ;\qquad \left. A_M\right|_{\theta=0}=  W_M
\ ;\qquad \left. A_A\right|_{\theta=0}=  0\nonumber\\
&& \left. \Xi^M\right|_{\theta=0}=  \xi^M\ ;\qquad
 \left. \Xi^A\right|_{\theta=0}= \epsilon^{\u A}{\cal A}_{\u A}^{-1\, A}\ .
 \label{Eattheta0}
\end{eqnarray}
Note the choice of the spinor-spinor part of the supervielbein. The matrix
appearing there is
related to the Killing spinors of the solution, which in general can be written as
\begin{equation}
\epsilon^{\u A}(x) = \eta^A {\cal A}_A{}^{\u A}(x)\ .   \label{defcA}
\end{equation}
Here $\epsilon^{\u A}(x)$ is the supersymmetry parameter, and the equation
determining the Killing spinors it the requirement that the transformation of
the fermions should be zero in this background. This determines the local
$\epsilon^{\u A}(x)$ in terms of rigid spinor parameters $\eta^A$. The
solutions which we consider here, preserve as many rigid supersymmetries  as
there are local supersymmetries. The gauge \eqn{Eattheta0} has been called the
Killing spinor gauge.\cite{Killinggauge}
\par
One compares the transformations of the fields to obtain $E$ at order
$\theta$. Then comparing the algebra leads to $\Xi$ at order
$\theta$. Afterwards one can obtain expressions at order $\theta^2$,
...~. Equivalent to gauge completion is to solve the supergravity torsions and curvature
constraints to all orders in $\theta$, which can in principle be done also order by order
in $\theta$.  It should be clear that this is a tedious task for theories with high
numbers of supersymmetry.  Fortunately, for certain backgrounds there is a shortcut and
one can solve the supergravity constraints to all orders in $\theta$ in closed form.  The
conditions on the background are that the gravitino vanishes and that the forms and dilaton
are covariantly constant.  The near-horizon limit of the brane solutions
given in Table~\ref{tbl:confpbr} satisfy these conditions \cite{KRAJ}.
The supergravity superspace has been derived explicitly in \cite{C}
for the M-branes and has been shown to be completely equivalent to the coset
superspace results.
\par
This brings us to the second method to obtain the geometric data of the background,
i.e.~the `{\sl supercoset approach}'\cite{superCosets,superCosetsCompl}. One constructs the
geometric connections for a supercoset $G/H$. In a $G$-covariant construction one decomposes
the generators of the superalgebra into bosonic ($B_{\u a}$) and fermionic
($F_{\u A}$) ones. With this decomposition the covariant derivative becomes
\begin{equation}
{\cal D} = d +  L^{\u a} B_{\u a} + L^{\u A} F_{\u A}\,.
\end{equation}
$L^{\u a}$ and $L^{\u A}$ are the bosonic and fermionic Cartan-forms and are given in
terms of the supercoset representative $G(Z)$
\begin{equation}
G(Z)^{-1} d G(Z) = L^{\u a} B_{\u a} + L^{\u A} F_{\u A}\,.
\end{equation}
The nilpotency ${\cal D}^2=0$ leads to Maurer--Cartan equations and
determines $dL$. In the following, we restrict the coset representative to the form $G(Z) = g(X)
e^{\Theta(X)^{\u A} F_{\u A}}$, with $g(X)$ is a representative of the bosonic subspace,
and $\Theta^{\u A}(X)$ is a linear in $\theta^A$ with $X$-dependent coefficients.
This form will allow us to find a solution to
 the Maurer--Cartan equations to all orders in $\theta$ in closed form \cite{superCosetsCompl}.
The general solution involves a matrix
\begin{equation}
\left( {\cal M}^2\right)_{\u A}{}^{\u B}=\Theta^{\u C} f_{\u C\u A}{}^{\u a}
\Theta^{\u D} f_{\u D\u a}{}^{\u B}\ ,
\end{equation}
where $f$ are the structure constants of $G$. The general solution reads
in Killing spinor gauge, i.e. $\left.L_{A}{}^{\u A}\right|_{\theta=0} = {\cal A}_A{}^{\u A}$ and
$\Theta^{\u A}(X) = \theta^A {\cal A}_A{}^{\u A}(X)$,
\begin{equation}
L^{\u A} = \left(d\theta {\cal A} \frac {\sinh {\cal M}}{{\cal M}} \right)^{\u A}\,,\qquad
L^{\u a} = L_0^{\u a} + 2 \Theta^{\u A} f_{\u A\u B}^{\u a} \left(d\theta {\cal A}
\frac{\sinh^2 {\cal M}/2}{{\cal M}^2}\right)^{\u B}\,, \label{LfromCoset}
\end{equation}
where $L_0^{\u a}$ is independent of $\theta$ and $d\theta$.
\par
Then one goes about identifying vielbeine and spin connections by splitting
the generators into `coset generators' $(K_{\u M},\,K_{\u A})$ and stability
group generators,
which generate $H$ (We assume that the stability group is pure
bosonic.).  The Cartan-forms in the $K$ directions are the supervielbeins
$(E^{\u M},\,E^{\u A})$, and the super spin connections
are the Cartan-forms in the $H$-directions.  For maximally supersymmetric backgrounds the
fermionic generators are all contained in $K$. In the supercoset approach the superforms
${\cal F}$ are constructed by trial and error from the supervielbeins, by demanding that they are
closed.
\section{Supersymmetric world-volume theory}  \label{ss:susyth}
The world volume actions are again given by \eqn{WVactions}.
In the Dirac-Born-Infeld term we use the induced metric
\begin{equation}
g^{\rm ind}_{\mu\nu} = \left( \partial_\mu Z^{\Lambda}
E_\Lambda{}^{\u{M}}\right) \left( \partial_\nu Z^{\Sigma}
E_\Sigma{}^{\u{N}} \right) \eta_{\u {M}\u {N}} \ .
\end{equation}
Furthermore there are the superforms which define the Wess-Zumino term.
For most of this section we will
concentrate on the  the black hole solutions of $d=4$, $N=2$ supergravity leading to
superconformal mechanics. In that case there is the super two-form
related to the graviphoton
${\cal F} = dA(Z)$, and the Wess-Zumino term for pure electrically charged
particles is
\begin{equation}
S_{WZ}= q \int d\tau\, \dot Z^\Lambda A_\Lambda   \ .
\end{equation}
 E.g. in flat superspace we have in that case
\begin{eqnarray}
&& E_M{}^{\u{M}}= \delta_M{}^{\u{M}}\ ; \qquad
E_{\alpha i}{}^{\u{M}}=\ft12\left( \gamma^{\u{M}}
\theta_i\right) _\alpha\nonumber\\
&&  A_M=0\ ;\qquad A_{\alpha i} =
\ft 12\varepsilon_{ij} \theta^j_\alpha \ .
\end{eqnarray}
Remark that the general spinor index $A$ is here $A= (\alpha i)$,
where $\alpha$ is a 4-dimensional spinor index, and $i=1,2$ are the $N=2$ indices.
The value of the supervielbein corresponds to the second part of \eqn{LfromCoset}
with
\begin{equation}
L_0^{\u M}=dX^{\u M}\ ;\qquad {\cal M}=0\ ;\qquad {\cal A}=\unity \ ;\qquad
f_{(\alpha i)(\beta j)}{}^{\u M}=\delta_{ij}\left( {\cal C}\gamma^{\u M}
\right) _{\u\alpha\u\beta}\ .\label{fABbh}
\end{equation}
This leads to
\begin{eqnarray}
 g^{\rm ind}_{\tau\tau} &=&\left(  \dot X^{ M}-\ft12\bar
\theta\Gamma^{ M}\dot\theta\right)
 \left( \dot X^{ N}-\ft12\bar \theta\Gamma^{ N}\dot\theta\right)
\eta _{ M  N}\nonumber\\
 S_{WZ}&=&\ft q2 \int d\tau
\dot {\bar \theta}^i \varepsilon_{ij} \theta^j \ .
\end{eqnarray}
\par
The $adS$ solutions preserve
rigid supersymmetry. The Killing spinors (solutions of
$\delta\psi_\mu=0$ in this background) are
\begin{equation}
\epsilon^i= M\left( \frac{M}{r}\right) ^{1/2}\eta_-^i + \left( \frac{r}{M}\right) ^{1/2}
\left( \eta_+^i - t\gamma_{0r} \eta_-^i\right)\ ,
\end{equation}
where spinors are split in $2\times 4$ real spinors as
\begin{equation}
\eta_\pm^i={\cal P}_{\pm}\eta^i=\frac12 \left( \eta^i \pm
\frac{1}{M}\left( Q+i\gamma_5 P\right)
\varepsilon^{ij}\gamma_0 \eta^j\right)\ ,
\end{equation}
$\eta_\pm^i$ depend only on $X^{\hat m}=(X^\theta,X^\phi)$ and
are Killing spinors of sphere
\begin{equation}
\nabla_{\hat m} \eta_{\pm}^i(\theta,\phi) =
\pm\frac{1}{2M}\gamma_r\gamma_{\hat m}  \eta_{\pm}^i (\theta,\phi) \end{equation}
There are 4 solutions for each sign \cite{LPR}. The commutators of these give
the two translations and rotation in $adS_2$ (in $(t,r)$) and $SO(3)$
transformations of the 2-sphere $(\theta,\phi)$.
These killing spinors provide us with the matrix ${\cal A}$ defined in
\eqn{defcA}. The relevant supercoset to construct is
\begin{equation}
\frac {SU(1,1|2)}{U(1) \times U(1)}\,
\end{equation}
and yields the supergeometry.
\par
Apart from rigid supersymmetries, i.e. the superisometries of
the background,  there is the counterpart
of world-volume diffeomorphisms, which is the (local) $\kappa$-symmetry.
It acts on the fermionic components of superspace as
\begin{equation}
\delta_\kappa\theta=(1+\Gamma)\kappa\ ,   \label{delthetakap}
\end{equation}
where $\Gamma$ is a complicated matrix in spinor space, function
of the world-volume fields, such that $\Gamma^2=\unity $ and $\Tr \Gamma=0$.
For a particular value,
e.g. on a classical solution, the operation in \eqn{delthetakap} is thus a
projection matrix. In the black hole case $\kappa$-symmetry requires
$q^2+p^2=m^2$ and
\begin{equation}
\Gamma=\frac 1{\sqrt{-g_{00}}}
 \varepsilon^{ij} \gamma_{\u M} E^{\u M}\ .
\end{equation}
At `classical values',
  $t=\tau$ and $r,\theta,\phi$ constant and vanishing fermions, and for
only electrically charged black hole ($P=0$ and $Q=M$)
\begin{equation}
\Gamma_{cl}=\varepsilon^{ij}\gamma_0\ ;\qquad  1+\Gamma_{cl}=2{\cal P}_-\ .
\end{equation}
$\kappa$ is a reducible symmetry, in the sense that if
$\kappa=(1-\Gamma)\kappa'$ in \eqn{delthetakap}, it does not contribute to the
transformations, i.e. $\kappa'$ is a zero mode. One can choose an
irreducible $\kappa$ symmetry: in our case, this is obtained if we
demand ${\cal P}_+\kappa=0$. E.g. for the M5 brane, the analogon
is a chiral projection on $\kappa$. Correspondingly
gauge fixing can be done by a chiral condition on $\theta$. Indeed
the expansion of \eqn{delthetakap} around the classical solution is
$\delta_\kappa \theta = {\cal P}_- \kappa + \ldots $, such that we
can gauge fix the $\kappa$ symmetry by imposing
\begin{equation}
{\cal P}_-\theta = 0\ .
\end{equation}
The remaining symmetry is determined by $(\delta_\kappa+\delta_\eta)
{\cal P}_-\theta =0$, which can be solved for $\kappa$ in terms of
$\eta_\pm$. This leads to a modification of the supersymmetry
transformations, similar to what happened in the bosonic case.
\par
The result is a conformal supersymmetry: where $\eta_+$ takes role of
$Q$-supersymmetry and $\eta_-$ of $S$-supersymmetry.
A similar mechanism works also for other brane backgrounds and
world-volumes. At the end the super-$adS_{d+1}$ is thus deformed to a
super$Conf_{d}$.

\section{Linearised action with superconformal symmetry}
\label{ss:lin}
The world--volume action can also be considered in a flat background.
As  above, after gauge fixing $X^m=\sigma^m$, the remaining bosonic
fields are $X^{m'}$ and possibly gauge fields. In line with the theme of this
workshop, we take as an example the $M5$
theory, arising from $D=11$ dimensions. The remaining scalars are thus the coordinates
of the 5 dimensions perpendicular to the brane, and the corresponding rotations are
$SO(5)=Sp(4)$, which will be the $R$-symmetry group. There is further an
antisymmetric tensor $B_{\mu \nu}$ on the world--volume.
On the fermionic side
half of the fermionic superspace coordinates disappear in the $\kappa$
gauge fixing, the other half remains as fermionic fields on the world-volume.
In the $M5$ example we start from 11 dimensions, with a 32 component spinor
$\theta$. This is split in chiral and anti-chiral spinors in 6
dimensions. One chirality is put to zero by the gauge condition, and the other
chirality are $4\times 4$ spinors $\lambda_i$:
4-components spinors of the 6-dimensional
theory, in a 4 of $SO(5)=Sp(4)$. This gives the content of a $(0,2)$ tensor
multiplet in 6 dimensions, and we have used again the auxiliary
field $a$ of the PST formulation as in \eqn{HwithPST}.
If we linearise the action in $X^{m'}$,
$B_{\mu \nu}$ and $\lambda_i$,
\begin{equation}
S_{lin}=\int d^6x\, \left[-\ft12 H^-_{\mu\nu}H^{*\mu\nu}
- \ft12 \partial_\mu X^{m'}\cdot \partial^\mu X^{m'}
+2\bar \lambda\dsl\lambda\right]\ ,\end{equation}
it turns out that there is again
conformal symmetry. We find ordinary supersymmetries with
left-handed parameter $\epsilon$ and special supersymmetries with
right-handed $\eta$
\begin{eqnarray*}
\delta X^{m'}&=&-2\bar \epsilon(x)\gamma^{m'}\lambda \ ,\nonumber\\
\delta\lambda&=& \ft12(\dsl X^{m'})
\gamma_{m'}\epsilon(x)-\ft16
h^+_{\mu\nu\rho}\gamma^{\mu\nu\rho}\epsilon(x) +2X^{m'}\gamma_{m'}
\eta \nonumber\\
\delta B_{\mu\nu}&=&-2\bar \epsilon(x)\gamma_{\mu\nu}\lambda
\ ;\qquad \delta a=0\nonumber\\
h_{\mu\nu\rho}^+&\equiv&\ft14 H_{\mu\nu\rho}
-\ft32 v_{[\mu}H^-_{\nu\rho]}\ ,
\end{eqnarray*}
with $\epsilon(x)=\epsilon+\gamma_\mu x^\mu\eta$.
The fields have Weyl weights
\begin{equation} w(X^{m'})=w(B_{\mu\nu})=2\ ;\qquad w(a)=0\ ;\qquad
w(\lambda)=\frac{5}{2}\ .
\end{equation}
The first one is in accordance with Table~\ref{tbl:confpbr}.
The algebra is the $N=4$ superconformal algebra
$OSp(8^*|4)$ as in Table~\ref{tbl:sca}.
\par
This occurence of the superconformal symmetry can be understood from
expanding the one in $adS$ background. The latter is for
small $R$ (or big $r$):
\begin{eqnarray}
S_{\rm M5}&=& S_{lin} + {\cal O}\left( \frac Rr \right)^3\left(
\mbox{ higher derivative terms }\right)\ .
\end{eqnarray}
\par
This mechanism works as well e.g. for
$M2$, where it gives the $N=8$ conformal scalar multiplet in 3 dimensions,
as for $D3$ where it leads to the $N=4$ conformal vector multiplet in 4 dimensions.


\section{Conclusions}
\label{ss:conclusions}
We spelled out the procedure \cite{conffads} establishing superconformal symmetry
of the gauge-fixed brane actions, starting from a
background with $adS_{d+1} \times S^{\tilde d+1}$ geometry.
The symmetries appear in the classical brane
actions before gauge-fixing local diffeomorphism and $\kappa$-symmetry,
and in the process of gauge-fixing of the latter, these rigid symmetries
become superconformal transformations.
\par
This symmetry also appears in a flat background, i.e.
$ISO(D-1,1)$ invariant, after gauge fixing and linearising, which is equivalent to
the leading terms of an expansion in
$R/r$ or dropping higher derivative terms of the action in $adS$ background.
For $M5$ one obtains a superconformal tensor multiplet in 6 dimensions,
including a conformal realization of the PST action. \cite{m5tens}
\par
We applied \cite{bhsconfm} the same procedure in $adS$ background to generalize the
`non-relativistic' superconformal mechanics of
Akulov and Pashnev, and Fubini and Rabinovici \cite{AP}, to a
 `{\it relativistic superconformal mechanics'}, having the former as
limit $M\rightarrow \infty$.

\section*{Acknowledgments}
The work of R. K. is supported by the NSF grant PHY-9870115. A.V.P.
thanks his employer, the F.W.O., Belgium, for the financial support.
Work supported by the European Commission TMR programme
ERBFMRX-CT96-0045.


\section*{References}
\begin{thebibliography}{99}
\bibitem{m5tens}P. Claus, R. Kallosh and A. Van Proeyen,
\Journal{\NPB}{518}{117}{1998}; hep-th/9711161.
\bibitem{conffads}P. Claus, R. Kallosh, J. Kumar, P.K. Townsend and
A. Van Proeyen, \Journal{\em JHEP}{06}{004}{1998} ; hep-th/9801206.
\bibitem{bhsconfm}P. Claus, M. Derix, R. Kallosh, J. Kumar, P.K. Townsend,
\Journal{\PRL}{81}{4553}{1998}; hep-th/9804177.
\bibitem{nahm}
W. Nahm, 
\Journal{\NPB}{135}{149} {1978}.
\bibitem{LieSA}
W. Nahm, V. Rittenberg and M. Scheunert,
\Journal{\em J. Math. Phys.}{17}{1626, 1640}{1976};\\
V.G. Kac,
\Journal{\em Commun. Math. Phys.}{53}{31}{1977}.
\bibitem{Annecy}
L. Frappat, P. Sorba and A. Sciarrino, {\it Dictionary on Lie Superalgebras}, hep-th/9607161.
\bibitem{realLieSA}
V.G. Kac,
\Journal{\em Adv. Math.}{26}{8}{1977}; \\
M. Parker,
\Journal{\em J. Math. Phys.}{21}{689}{1980}.
\bibitem{classinf} P. Ramond and J.H. Schwarz,
\Journal{\PLB}{64}{75}{1976};\\
A. Sevrin, W. Troost and A. Van Proeyen,
\Journal{\PLB} {208}{447}{1988}.
\bibitem{JWvHAVP} J.W. van Holten and A. Van Proeyen,
\Journal{{\em J. Phys.}A}{15}{3763}{1982}.
\bibitem{Mfromsa}
P.K. Townsend, 
   hep-th/9712004.
\bibitem{GT} G. W. Gibbons and P. K. Townsend,
\Journal{\PRL}{71}{3754}{1993}; hep-th/9307049.
\bibitem{Maldacena} J. Maldacena,
\Journal{\em Adv. Theor. Math. Phys.}{2}{231}{1998}; hep-th/9711200.
\bibitem{specialdual} S. Hyun,
hep-th/9704005;\\
H.J.~Boonstra, B.~Peeters and K.~Skenderis,
\Journal{\PLB}{411}{59}{1997}; hep-th/9706192;\\
E. Cremmer, I.V. Lavrinenko, H. Lu, C.N. Pope, K.S. Stelle and T.A. Tran,
\Journal{\NPB}{534}{40}{1998}; hep-th/9803259.
\bibitem{GMT}
 J. P. Gauntlett, R. C. Myers, P. K. Townsend,  hep-th/9810204.
\bibitem{KPSk2}
H.J.~Boonstra, B.~Peeters and K.~Skenderis,
\Journal{\NPB}{533}{127}{1998}; hep-th/9803231;\\
S. Elitzur, O. Feinerman, A. Giveon, D. Tsabar,
hep-th/9811245.
\bibitem{PST} P. Pasti, D. Sorokin and M. Tonin,
\Journal{\PLB}{398}{41}{1997}; hep-th/9701037.
\bibitem{BrownHenn}
J.D. Brown and M. Henneaux,
\Journal{\em Commun. Math. Phys.}{104}{207}{1986}.
\bibitem{BGS} F. Brandt, J. Gomis and J. Sim\'on,
\Journal{\PLB}{419}{148}{1998}, hep-th/9707063; and \Journal{\PRL}{81}{1770}{1998};
hep-th/9803196.
\bibitem{DFF} V.~de~Alfaro, S.~Fubini and G.~Furlan,
\Journal{\NCA}{34}{569}{1976}.
\bibitem{AP}
V.P.~Akulov and I.A.~Pashnev,
\Journal{\em Theor.~Math.~Phys.}{56}{862}{1983};\\
S.~Fubini and E.~Rabinovici,
\Journal{\NPB}{245}{17}{1984}.
\bibitem{Gaugecompl}
P. Nath and  R. Arnowitt, \Journal\PLB{65}{73}{1976};\\
E. Cremmer and  S. Ferrara, \Journal\PLB{91}{61}{1980};\\
L. Castellani, P. van Nieuwenhuizen and  J. Gates,
\Journal\PRD{22}{2364}{1980};\\
B. de Wit, K. Peeters and J. Plefka,
\Journal{\NPB}{532}{99}{1998}; hep-th/9803209
\bibitem{Killinggauge}
R. Kallosh,
hep-th/9807206.
\bibitem{KRAJ}
R. Kallosh and A. Rajaraman,
hep-th/9805041.
\bibitem{C}
P. Claus,
hep-th/9809045.
\bibitem{superCosets}
L. Castellani, A. Ceresole, R. D'Auria, S. Ferrara, P. Fr\`e and M. Trigiante,
\Journal{\NPB}{527}{142}{1998}; hep-th/9803039\\
R.R. Metsaev and A.A. Tseytlin,
\Journal{\NPB}{533}{109}{1998};
hep-th/9805028 and
\Journal{\PLB}{436}{281}{1998};  hep-th/9806095
\bibitem{superCosetsCompl}
R. Kallosh, J. Rahmfeld and A. Rajaraman,
\Journal{\em JHEP}{09}{002}{1998};
hep-th/9805217.\\
B. de Wit, K. Peeters, J. Plefka and A. Sevrin,
%
hep-th/9808052.
\bibitem{LPR}
H. L\"u, C.N. Pope and J. Rahmfeld,
hep-th/9805151.
\end{thebibliography}
\end{document}